\documentclass[aip,cha, reprint,showpacs,showkeys]{revtex4-1}
\usepackage{amssymb, dcolumn,  graphicx, latexsym, amsfonts, bm}

\begin{document}

\title{Ion-temperature-gradient sensitivity of the hydrodynamic instability caused
by shear in the magnetic-field-aligned plasma flow}
\author{V. V. Mikhailenko}\email[E-mail: ]{vladimir@pusan.ac.kr}
\affiliation{Pusan National University,  Busan 609--735, S. Korea.}
\author{V. S. Mikhailenko}
\affiliation{V.N. Karazin Kharkiv National University, 61108 Kharkiv, Ukraine.}
\affiliation{Kharkiv National Automobile and Highway University, 61002 Kharkiv,
Ukraine.}
\author{Hae June Lee}
\affiliation{Pusan National University, Busan 609--735, S. Korea.}
\author{M.E.Koepke}
\affiliation{Department of Physics, West Virginia University, 26506 Morgantown, WV, USA}

\begin{abstract}
The cross-magnetic-field (i.e., perpendicular) profile of ion temperature and the
perpendicular profile of the magnetic-field-aligned (parallel) plasma flow are
sometimes inhomogeneous for space and laboratory plasma. Instability caused by a
gradient in either the ion-temperature profile or by shear in the parallel flow has
been discussed extensively in the literature. In this paper, hydrodynamic plasma
stability is investigated, real and imaginary frequency are quantified over a range
of the shear parameter, the normalized wavenumber, and the ratio of density-gradient
and ion-temperature-gradient scale lengths, and the role of inverse Landau damping
is illustrated for the case of combined ion-temperature gradient and parallel-flow
shear. We find that increasing the ion-temperature gradient reduces the instability
threshold for the hydrodynamic parallel-flow shear instability, also known as the
parallel Kelvin-Helmholtz 
instability or the D'Angelo instability. We also find that a kinetic instability
arises from the coupled, reinforcing action of both free-energy sources. For the
case of comparable electron and ion temperature, we illustrate analytically the
transition of the D'Angelo instability to the kinetic instability as the shear
parameter, the normalized wavenumber, and the ratio of density-gradient and
ion-temperature-gradient scale lengths are varied and we attribute the changes in
stability to changes in the amount of inverse ion Landau damping. We show that, near
a normalized wavenumber $k_{\perp} \rho_{i}$ of order unity, the real and imaginary
values of frequency become comparable and the growth rate maximizes.  
\end{abstract}

\maketitle


\section{INTRODUCTION}\label{sec1}
Shear in magnetic-field-aligned (i.e., parallel) plasma flow can be found
in space\cite{Amatucci, Kintner, McFadden}, fusion\cite{Asakura, LaBombard, 
Fedorczak, Pedrosa, Wang} and laboratory\cite{Agrimson1, Koepke2002, Koepke2003,
Koepke2004, Koepke2006, Sen,  Kaneko-03, Kaneko-05} plasma. Ionospheric regions of
inhomogeneous parallel plasma flow can map magnetically to magnetospheric regions of
large-scale field-aligned currents, resulting in broadband electrostatic noise in
the auroral zone of geospace \cite{Amatucci}. 

Experimental observations from a number tokamaks and stellarators\cite{Asakura, 
LaBombard, Fedorczak, Pedrosa, Wang} have found large nearly sonic parallel sheared flows 
inside the last closed flux surface. Near-sonic parallel sheared flows are systematically 
observed in the far scrape-off layer (SOL) of the X-point divertor tokamaks 
JT-60\cite{Asakura} and Alcator C--Mod \cite{LaBombard} tokamaks, in the limiter tokamak 
Tore Supra\cite{Fedorczak}. These flows are deemed unstable\cite{McCarthy, Garbet, 
Schwander} against the development of the well known hydrodynamic D'Angelo instability
\cite{D'Angelo}. Schwander et al.\cite{Schwander} showed that plasma might be unstable to the 
parallel-flow shear instability 
around limiters, as inferred from the experimental findings of Fenzi et al.\cite{Fenzi}, 
thereby explaining local enhancements of turbulence and showing that,
according to the local linear stability criterion, stability is sensitive to core
parallel rotation. Their work speaks to the interest that would be given to future numerical
modelling of experimentally relevant plasma conditions to assess both the properties
of the transport coefficients associated with the parallel-flow shear-driven
instability and the presence of free-energy that would support turbulence that
arises from that instability. In this paper, using analytical theory and numerical
analysis, we quantify the real and imaginary frequency of the parallel-flow
shear-driven instability over a range of the shear parameter, the normalized
wavenumber, and the ratio of density-gradient and ion-temperature-gradient scale
lengths and demonstrate that the flow-shear threshold for instability reduces as the
ion-temperature gradient increases. We also illustrate the role of inverse Landau
damping in this case of combined ion-temperature gradient and parallel-flow shear. 

If the parallel-ion-flow shear is accompanied by inhomogeneous ion 
temperature, a hydrodynamic instability can develop into a kinetic instability as
will be shown. The investigation of plasma stability in the presence of these two
free-energy sources
was initiated by S.Migliuolo in his investigations of the plasma sheet 
boundary layer\cite{Migliuolo} in the Earth’s magnetosphere. He found, that kinetic
instability arises by 
virtue of the coupled action of both parallel-velocity shear $V'_{0}(x)$ and an ion 
temperature gradient that reinforce each other. 
Inverse ion Landau damping is responsible for the combined ion-temperature-gradient
flow-shear-driven (ITG-FSD) kinetic instability, a conclusion quite different from
D'Angelo’s conclusion for the homogeneous-temperature, hydrodynamic FSD instability
in the framework of the two-fluid equations. The calculations in Ref.
\cite{Migliuolo} involved taking the large argument ($z_{i}=
\omega/\sqrt{2}k_{z}v_{Ti}\gg 
1$) limit of the ion plasma dispersion function and examining marginal stability,
without quantifying the maximum growth rate. The linear analysis of the
parallel-flow shear stability in the presence of 
inhomogeneous ion temperature with application to the edge tokamak plasma was
performed in 
Ref.\cite{Rogister}. It was predicted, that the roles of magnetic shear, trapped
electrons, 
and toroidal curvature are negligible for the  ITG--SFD kinetic instability and the
behavior of  the unstable growth rate was calculated only in the neighborhood of
marginal instability.

In this paper, we present results from a numerical and analytical investigation of the 
sensitivity of the hydrodynamic D'Angelo mode to the ion temperature gradient 
and we interpret the transition of the hydrodynamic instability to the ITG--SFD
kinetic instability.  Extending beyond the result for marginal stability of
Ref.\cite{Rogister}, we arrive at the approximate analytical solution of the
dispersion relation for the parameters associated with the maximum value of the
growth rate.

The paper is organized as follows: The basic equations are presented in
Sec.\ref{sec2}. In 
Section \ref{sec3}, we analyze the effect of the finite ion temperature and ion
temperature 
gradient on the hydrodynamic D'Angelo mode. In Section \ref{sec4}, we consider the
ITG--SFD 
kinetic instability.  The Conclusions are presented in Sec.\ref{sec5}.

\section{BASIC EQUATIONS}\label{sec2}

We consider a kinetic Vlasov-Poisson model of inhomogeneous, magnetic-field-aligned,
single-ion-species, plasma flow with velocity ${\bf{V}}_{0}(X_{\alpha})\parallel to
B_{0} {\bf{e}}_z $. 
The Vlasov equation for the perturbation $f_{\alpha}=F_{\alpha}-F_{0\alpha}$ 
of the distribution function $F_{\alpha}$ with equilibrium function $F_{0\alpha}$ 
in  guiding center coordinates in slab geometry,
$X_{\alpha}=x+\frac{v_{\bot}}{\omega_{c
\alpha}}\sin\phi$, $Y_{\alpha}=y-\frac{v_{\bot}}{\omega_{c\alpha}}\cos\phi$,  where
$\omega 
_{c\alpha}$ is the cyclotron frequency, has a form
\begin{eqnarray}
& \displaystyle \frac{\partial f_{\alpha}}{\partial t}-\omega_{c}
\frac{\partial f_{\alpha}}{\partial\phi} +v_{z}\frac{\partial
f_{\alpha}}{\partial z} \nonumber
\\ & \displaystyle =
\frac{e}{m\omega_{c}}\frac{\partial\Phi}{\partial Y}
\frac{\partial F_{0\alpha}}{\partial X}-\frac{e}{m}\frac{\omega_{c}}{v_{\bot}}
\frac{\partial\Phi}{\partial \phi} \frac{\partial F
_{0\alpha}}{\partial
v_{\bot}} +\frac{e}{m}\frac{\partial\Phi}{\partial z}
\frac{\partial F_{0\alpha}}{\partial v_{z}}.
\label{1}
\end{eqnarray}

In what follows, $F_{i0}$ is considered as the shifted Maxwellian 
distribution function for electrons and ions ($\alpha=i,e$)
\begin{eqnarray}
& \displaystyle F_{0\alpha} = \frac{n_{0\alpha}\left(X_{\alpha}
\right)}{\left({2\pi v_{T\alpha}^2}\right)^{3 /2}}\exp\left[-
\frac{v_{\perp}^{2}}{2v_{T\alpha}^2} - \frac{\left(v_{z} -
V_{0}(X_{\alpha})\right)^2}{2v_{T\alpha}^{2}} \right],
\label{2}
\end{eqnarray}
assuming the inhomogeneity direction of the density and temperature of the
sheared-flow 
species is along coordinate $X_{\alpha}$, $v_{T\alpha}=\left(T_{\alpha}\left(X_{\alpha}
\right)/m_{\alpha} \right)^{1/2}$ is the thermal velocity. The flow velocity of ions $
\mathbf{V}_{0}$ is assumed to be equal to that of the electrons which is consistent
with the fluid approximation used
for the Kelvin-Helmholtz instability but inconsistent with including the development
of current-driven 
instabilities. In order to simplify the problem, a 
velocity $\mathbf{v}$ usually transforms from the laboratory to a convecting frame
of reference, 
$\mathbf{v}= \hat{\mathbf{v}}+V_{0}(X_{\alpha})
\mathbf{e}_{z}$. We consider here the idealized inhomogeneous-flow case of 
homogeneous parallel-velocity shear, i.e. $V'_{0}=const$. After transformation, the
Vlasov 
equation takes the form 
\begin{eqnarray}
& \displaystyle \frac{\partial f_{\alpha}}{\partial t}-\omega_{c}
\frac{\partial f_{\alpha}}{\partial\phi} +\left(\hat{v}_{z}+V_{0}(X_{\alpha}) \right) 
\frac{\partial
f_{\alpha}}{\partial z} \nonumber
\\ & \displaystyle 
=\frac{e}{m\omega_{c}}\frac{\partial\Phi}{\partial Y}\left(\frac{\partial F_{0\alpha}}
{\partial X}-V'_{0}\frac{\partial F_{0\alpha}}
{\partial \hat{v}_{z}}\right) 
\nonumber
\\ & \displaystyle -\frac{e}{m}\frac{\omega_{c}}{v_{\bot}}
\frac{\partial\Phi}{\partial \phi} \frac{\partial F_{0\alpha}}{\partial
v_{\bot}} +\frac{e}{m}\frac{\partial\Phi}{\partial z}
\frac{\partial F_{0\alpha}}{\partial v_{z}}.
\label{3}
\end{eqnarray}
The general approach to solving Eq.(\ref{3}) is a Fourier transform over time and
space coordinates employing the local approximation, for which restrictions $k_x
L_{n} \gg 1$, $k_xL_{T_{i}} \gg 1$ and $k_x L_v \gg 1$ are assumed, where $L_{n} =
\left[ d
\ln n_{0} \left(X \right)/dx \right]^{-1}$, $L_{T_{i}} = \left[ d\ln T_{i}
\left(X \right)/dx \right]^{-1}$ and $L_{v}= \left[ d\ln V_0
\left(X\right)/dx \right]^{-1}$. Although the local approximation is typically
justifiable in the case of inverse electron Landau damping in homogeneous plasma,
justification of the local approximation in the case of velocity shear $V'_{0}$
having a value comparable to the ion cyclotron frequency requires careful and
convincing arguments.

 After the Fourier transformation over the space coordinates, Eq.(\ref{3}) becomes
\begin{eqnarray}
& \displaystyle \frac{\partial f_{\alpha}}{\partial t}-\omega_{c}
\frac{\partial f_{\alpha}}{\partial\phi} +ik_{z}\hat{v}_{z}f_{\alpha}\left(v_{\bot},
\phi, 
\mathbf{k}, t\right)
-V'_{0}k_{z}\frac{\partial f_{\alpha}}
{\partial k_{x}} \nonumber
\\ & \displaystyle 
=\frac{e}{m\omega_{c}}ik_{y}\Phi\left(\mathbf{k}, t\right)\left(\frac{\partial
F_{0\alpha}}
{\partial X}-V'_{0}\frac{\partial F_{0\alpha}}
{\partial \hat{v}_{z}}\right) 
\nonumber
\\ & \displaystyle -\frac{e}{m}\frac{\omega_{c}}{v_{\bot}}
\frac{\partial\Phi}{\partial \phi} \frac{\partial F
_{0\alpha}}{\partial
v_{\bot}} +i\frac{e}{m}k_{z}\Phi
\frac{\partial F_{0\alpha}}{\partial v_{z}}.
\label{4}
\end{eqnarray}
where any spatially homogeneous part of flow velocity is eliminated from the problem
by a simple Galilean transformation. In deriving from Eq.(\ref{4}) the equation that
couples $f_{\alpha}$ with potential $\varphi$  of each separate spatial Fourier mode, 
we have to exclude from Eq.(\ref{4}) the term $-V'_{0}k_{z}\frac{\partial 
f_{\alpha}}{\partial k_{x}}$, due to which the Fourier modes of $f_{\alpha}$ appear
to be coupled with all Fourier modes of the electrostatic potential $\varphi$. The
characteristic equation 
\begin{eqnarray}
&\displaystyle
dt=-d\hat{k}_{x}/V'_{0}\hat{k}_{z}
\label{5} 
\end{eqnarray}
gives the solution $k_{x}+V'_{0}tk_{z}=K_{x}$, where $K_{x}$ 
as the integral of Eq.(\ref{5}) is time independent. It reveals that $f_{\alpha}
=f_{\alpha}\left(K_{x}, k_{y}, k_{z}, t\right)=f_{\alpha}\left(k_{x}
+V'_{0}tk_{z}, k_{y}, k_{z}, t\right)$, 
i.e. the wave number components $k_{x}$ and $k_{z}$ have to be changed 
in such a way that $k_{x}+V'_{0}tk_{z}$ is left unchanged with time. For the 
instabilities considered in this paper $k_{x}\gg k_{z}$, the solution of
Eq.(\ref{4}) in the 
convecting frame of reference is of the modal form during a long time until
$V'_{0}t\lesssim 
k_{x}/k_{z}$. Until that time, the general dispersion equation is valid in the local 
approximation and is given in this model by
\begin{eqnarray}
& \displaystyle
1+k^2\lambda _{Di }^2
+ i\sqrt {\frac{\pi }{2}}
\frac{\left(\omega - k_{y} v_{di}\left(1-\frac{1}{2}\eta_{i} \right)
\right)}{k_{z} v_{Ti} } \nonumber \\  &\displaystyle 
\times\sum\limits_{n = - \infty }^{\infty}
W\left(z_{in}\right) A_{in} \left(k_{\bot} ^{2} \rho
_{i}^{2 } \right)
\nonumber \\  &\displaystyle
-\frac{k_{y} }{k_{z} }\frac{V'_{0 }}{\omega_{ci}}\left[1 + i\sqrt{\pi}
\sum\limits_{n =-
\infty }^{\infty} z_{i n} W\left(z_{i n} \right)
A_{i n} \left(k_{\bot}^{2}\rho_{i}^{2}\right)\right]
\nonumber
\\
&\displaystyle
-\eta_{i}\chi_{i}\sum\limits_{n =
- \infty }^{\infty}z_{in}\left(1+i\sqrt{\pi}z_{in}W\left(z_{i n}\right) \right)A_{i n} 
\left(k_{\bot} ^{2} \rho_{i}^{2 } \right)
\nonumber
\\
&\displaystyle
-\eta_{i}\chi_{i}
\sum\limits_{n = - \infty }^{\infty}i\sqrt{\pi}W\left(z_{i n} \right)e^{-k_{\bot}^{2}
\rho_{i}^{2}}
\nonumber \\  &\displaystyle \times
k_{\bot}^{2}\rho_{i}^{2}\left[I_{n}\left(k_{\bot}^{2}\rho_{i}^{2}\right)-I'_{n}
\left(k_{\bot}^{2}\rho_{i}^{2}\right) \right]
\nonumber
\\
&\displaystyle
+\frac{T_{i}}{T_{e}} \left(1+i\sqrt{\frac{\pi}{2}}\frac{\left(\omega - k_{y} v_{de}
\right)}{k_{z} v_{Te} }W\left(z_{e} \right)\right)=0,
\label{6}
\end{eqnarray}
In Eq.(\ref{6}),  $\lambda_{Di}$  is the ion Debye length, $\omega_{ci}$
and $\rho_{i}= v_{Ti}/\omega_{ci}$  is the ion thermal Larmor radius,
$A_{n}\left(k_\bot^2 
\rho_{i}^2\right)=I_{n}\left(k_\bot^2 \rho_{i}^2\right)e^{-k_\bot^2 \rho_{i}^2}$, 
$I_{n}$ is 
the modified Bessel function of order $n$, $z_{i n} = \left(\omega-n\omega_{ci}\right)/
\sqrt{2}k_{z}v_{Ti}$, $z_{e} = \omega/\sqrt{2}k_{z}v_{Te}$, $\chi_{\alpha} = k_{y}v_{d
\alpha}/\sqrt{2}k_{z}v_{Ti}$,
$v_{d\alpha}=\left(v_{T\alpha}^{2}/\omega_{c\alpha}\right)
\left(d\ln n_{0}\left(x\right)/dx\right)$ is the diamagnetic 
drift velocity of ions $\left( \alpha=i\right) $, and electrons
$\left(\alpha=e\right)$, $
\eta_{i}= d\ln T_{i}/d\ln n_{i}$ which is approximately $L_{T_{i}}$/$L_{n_{i}}$,
$W\left(z\right)=e^{ - z^{2}}\left(1 +\left(2i / \sqrt {\pi 
}\right)\int\limits_{0}^{z} e^{t^{2}}dt \right)$ is the complex error function. 
The principal difference of the Eq.(\ref{6}) with similar dispersion equation, 
obtained in local approximation under condition $k_x L_v \gg 1$
of the "slow spatial variation" of the velocity $V_{0} (x)$,
is that the frequency $\omega$  does not contain any more a spatially inhomogeneous
part of 
the Doppler shift $k_{z}V'_{0}x$, which may be safely omitted during the long time
$t\lesssim 
\left(k_{x}/V'_{0}k_{z}\right)$ when the convective coordinates are used.

We consider low frequency modes with frequency $\omega$ much less then the ion
cyclotron 
frequency $\omega_{ci}$ in the limit $|\omega|\lesssim k_{z}v_{Te}$ as is
appropriate for 
velocity-shear and temperature-gradient instabilities. For these conditions, the
general 
dispersion equation that accounts for parallel-flow shear and inhomogeneous profiles
of both 
ion density and ion temperature and that accounts for the effects of thermal motion
of ions, 
both along and across the magnetic field, is \cite{Migliuolo}
\begin{eqnarray}
& \displaystyle 1+\frac{T_{i}}{T_{e}}\left(1+\Delta\varepsilon_{e}\left(\mathbf{k},
\omega\right)\right)-\left(\frac{k_{y}V'_{0}}{k_{z}\omega_{ci}}+ z_{i}\eta_{i}\chi_{i}\right)A_{0i} \left(k_{\bot}^{2}\rho_{i}^{2}\right)
\nonumber
\\
&\displaystyle
+i\sqrt{\pi}W\left(z_{i}\right)
\left\{\left[z_{i}\left(1-\frac{k_{y}V'_{0}}{k_{z}\omega_{ci}}  \right)
\right.\right.\nonumber \\  &\displaystyle
\left.\left.
-\chi_{i}\left(1- \frac{\eta_{i}}{2}\left( 1-2z_{i}^{2}\right) \right)\right]A_{0i}
\left(k_{\bot}^{2}\rho_{i}^{2}\right)\right.
\nonumber
\\
&\displaystyle
\left.+\chi_{i}\eta_{i}k_{\bot}^{2}\rho_{i}^{2}\left(A_{0i}\left(k_{\bot}^{2}\rho_{i}^{2}
\right)-
A_{1i}\left(k_{\bot}^{2}\rho_{i}^{2}\right)\right) \right\}=0,
\label{7}
\end{eqnarray}
where $z_{i} = \omega/\sqrt{2}k_{z}v_{Ti}$, $\Delta \varepsilon_{e}\left(\mathbf{k},
\omega\right)=i\sqrt{\pi}\, W\left(z_{e}\right) $ $\times\left(z_{e}-\chi_{e}\right)$. 
The goal of this paper is the numerical and analytical investigation of Eq.(\ref{7})
for the conditions at which thermal effects of ions having an inhomogeneous profile
are dominant.

\section{Hydrodynamic ions}\label{sec3}
In the long-parallel-wavelength limit, $|z_{i}|\gg 1$, in which ion Landau 
damping is negligible, the dispersion equation (\ref{7}) reduces to the form
\begin{eqnarray}
&\displaystyle
1+\frac{T_{i}}{T_{e}}\left(1+\Delta\varepsilon_{e}\left(\mathbf{k}, \omega
\right)\right)-
A_{0i}\left(k_{\bot}^{2}\rho_{i}^{2}\right)
\nonumber \\  &\displaystyle
-\left(1-\frac{k_{y}V'_{0}}{k_{z}\omega_{ci}}  
\right)\frac{k^{2}_{z}v^{2}_{Ti}}{\omega^{2}}A_{0i}\left(k_{\bot}^{2}\rho_{i}^{2}\right)
\nonumber \\  &\displaystyle
+\frac{k_{y}v_{di}}{\omega}\left(1+\left(1+\eta_{i} \right) \frac{k^{2}_{z}v^{2}_{Ti}}
{\omega^{2}} \right) A_{0i}\left(k_{\bot}^{2}\rho_{i}^{2}\right)
\nonumber \\  &\displaystyle
-\eta_{i}\frac{k_{y}v_{di}}{\omega}\left(1+\frac{k^{2}_{z}v^{2}_{Ti}}{\omega^{2}} 
\right)
\nonumber \\  &\displaystyle \times
k_{\bot}^{2}\rho_{i}^{2}\left(A_{0i}\left(k_{\bot}^{2}\rho_{i}^{2}\right)
-A_{1i}\left(k_{\bot}^{2}\rho_{i}^{2}\right)\right)=0.
\label{8}
\end{eqnarray}
For plasma without parallel-flow shear, this equation describes the hydrodynamic ion
temperature 
gradient drift instability and the kinetic ion temperature gradient 
drift instability developed due to the inverse electron Landau damping of the drift
waves when $k_{z}v_{Ti}\ll \omega \lesssim k_{z}v_{Te}$. In the presence of sheared plasma
flow, this equation in the case 
$1-k_{y}V'_{0}/k_{z}\omega_{ci}>0$ determines the shear-modified ion acoustic 
instability\cite{Gavrishchaka, Gavrishchaka1}, which can be excited due to the inverse 
electron Landau damping for wide range of ion-electron temperature ratios even for ion-
electron temperature ratios of the order of unity and lager. 
In this paper we consider the case with $k_{y}V'_{0}/k_{z}\omega_{ci}>0$ in which the 
hydrodynamic D'Angelo instability\cite{D'Angelo} develops with frequency $\omega
\left(\mathbf{k} \right)$,
\begin{eqnarray}
&\displaystyle
\omega\left(\mathbf{k} \right)=-k_{y}v_{di}\frac{\left(A_{0i}-\eta_{i}
k_{\bot}^{2}\rho_{i}^{2}\left(A_{0i}-A_{1i} \right)  \right) }{2\left(1+\frac{T_{i}}{T_{e}}-A_{0i}\right)}
\label{9}
\end{eqnarray}
and with the growth rate $\gamma\left(\mathbf{k} \right)$,
\begin{eqnarray}
&\displaystyle
\gamma\left(\mathbf{k} \right)=\frac{1}{\left(1+\frac{T_{i}}{T_{e}}-A_{0i} \right)}
\left[\left(\frac{k_{y}V'_{0}}{k_{z}\omega_{ci}}-1 \right)\right.
\nonumber \\  &\displaystyle \times k^{2}_{z}v^{2}_{Ti} A_{0i}\left(1-
A_{0i}+ \frac{T_{i}}{T_{e}}\right)
\nonumber \\  &\displaystyle
\left.-\frac{1}{4}k^{2}
_{y}v^{2}_{di}B^{2}_{i}\left(\eta_{i}, k_{\bot}^{2}\rho_{i}^{2} \right)  
\right]^{1/2},
\label{10}
\end{eqnarray}
where
\begin{eqnarray}
&\displaystyle B_{i}\left(\eta_{i}, k_{\bot}^{2}\rho_{i}^{2}\right)
=A_{0i}\left(k_{\bot}^{2}\rho_{i}^{2}\right)
\nonumber \\  &\displaystyle
-\eta_{i}k_{\bot}^{2}\rho_{i}^{2}\left(A_{0i}\left(k_{\bot}^{2}\rho_{i}^{2}\right)
-A_{1i}\left(k_{\bot}^{2}\rho_{i}^{2}\right)\right),\label{11}
\end{eqnarray}
Eqs.(\ref{9}) and (\ref{10}) account for the thermal motion of ions across the
magnetic field.  We assume in what follows, that $|k_{y}V'_{0}/k_{z}\omega_{ci}|\ll
m_{i}/m_{e}$. Under that condition we have $|z_{e}|\ll 1$, and the term $\Delta
\varepsilon_{e}\left(\mathbf{k}, \omega \right)$, which determines the effect of electron 
Landau damping, may be neglected in Eq.(\ref{10}). The negative term 
containing the ion diamagnetic drift velocity in Eq.(\ref{10}) indicates that plasma
density 
inhomogeneity acts as a stabilizing factor for the hydrodynamic D'Angelo instability, 
whereas the $B_{i}\left(k_{\bot}^{2}\rho_{i}^{2}; \eta_{i}\right)$ term,
representing the effect of ion-temperature inhomogeneity, reduces the density gradient's 
stabilizing effect by reducing the magnitude of $B^{2}_{i}$ and reinforces the development of 
the hydrodynamic D'Angelo instability.

The D'Angelo instability with growth rate (\ref{10}) occurs for the values of $k_{z}$
bounded by the region $k_{z1}>k_{z}>k_{z2}$ in the $(k_{\bot}, k_{z})$ plane, where
$k_{z1,2}
$ are equal to
\begin{eqnarray}
&\displaystyle k_{z1,2}=\frac{k_{y}V'_{0}}{2\omega_{ci}}\left(1 \pm
\sqrt{1-\frac{v^{2}_{di}}
{\left( \rho_{i}V'_{0}\right)^{2}}G_{i}\left(k_{\bot}^{2}\rho_{i}^{2},
\eta_{i}\right)  
}\right), \label{12}
\end{eqnarray}
with
\begin{eqnarray}
&\displaystyle G_{i}\left(k_{\bot}^{2}\rho_{i}^{2}, \eta_{i}\right) =\frac{B^{2}_{i}
\left(k_{\bot}^{2}\rho_{i}^{2},
\eta_{i}\right)}{A_{0i}\left(k_{\bot}^{2}\rho_{i}^{2}\right)
\left(1+\frac{T_{i}}{T_{e}}-A_{0i}\left(k_{\bot}^{2}\rho_{i}^{2}\right) \right)}.
\label{13}
\end{eqnarray}
It follows from Eq.(\ref{12}), that such interval exists for velocity shearing rate
above the 
threshold value,  $V'_{0}>\omega_{ci}\left(\rho_{i}/L_{n}\right)\sqrt{ G_{i}}$.

\begin{figure}[htb]
\includegraphics[width=0.45\textwidth]{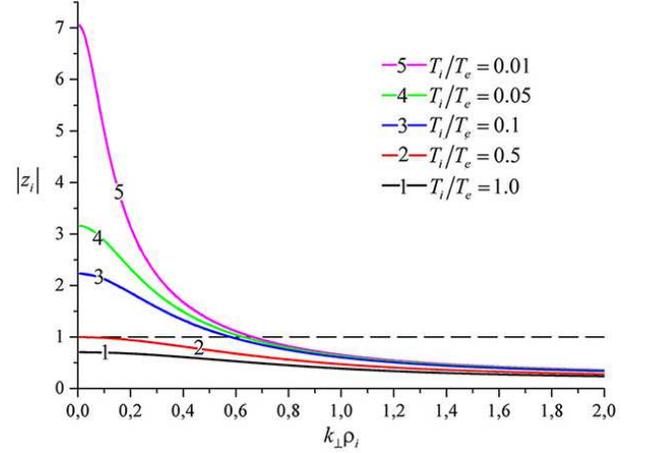}
\caption{\label{fig0} The argument $z_{i}$ of the plasma dispersion function for the
maximum
growth rate (14) of the D'Angelo hydrodynamic instability versus $k_{\bot}\rho_{i}$.
The 
dashed line corresponds to $z_{i}=1$. Below that line, Eq.(10) for the gydrodynamic
growth rate is not valid.}\end{figure}

The growth rate (\ref{10}), as a function of $k_{z}\rho_{i}$, attains its maximum
between 
$k_{z1}$ and $k_{z2}$, at $k_{z}=\left( V'_{0}/2\omega_{ci}\right)k_{y}$, and 
at values of $k_{y}$ for which the function $B_{i}\left(k_{\bot}^{2}\rho_{i}^{2},
\eta_{i}
\right)$ vanishes for the given value of $\eta_{i}$. The maximum growth rate is
equal to
\begin{eqnarray}
&\displaystyle \gamma\left(\mathbf{k}
\right)=\frac{k_{z}v_{Ti}A_{0i}^{1/2}\left(k_{\bot}^{2}
\rho_{i}^{2}\right)}
{\left(1+\frac{T_{i}}{T_{e}}-A_{0i}\left(k_{\bot}^{2}\rho_{i}^{2}\right)
\right)^{1/2}}.
\label{14}
\end{eqnarray}
The  hydrodynamic treatment for the D'Angelo instability is valid when the condition $|
z_{i}|>1$ holds for the whole interval $k_{z1}>k_{z}>k_{z2}$. Because the growth rate 
(\ref{10}) is greater than the frequency (\ref{9}), when D'Angelo instability
develops,  $|
z_{i}|$ may be expressed approximately as
\begin{eqnarray}
&\displaystyle |z_{i}| \approx\frac{\gamma}{\sqrt{2}k_{z1}v_{Ti}}
=\sqrt{ \frac{A_{0i}\left(k_{\bot}^{2}\rho_{i}^{2}\right)}{2\left( 1+\frac{T_{i}}
{T_{e}}-A_{0i}\left(k_{\bot}^{2}\rho_{i}^{2}\right)\right) } }.
\label{15}
\end{eqnarray}
It follows from Eq.(\ref{15}) that for comparable temperatures of the ions and
electrons, 
which is the case relevant to space, Q-machine, and tokamak plasmas, and/or for the 
perturbations of the order of the ion Larmor radius, $k_{\bot} \rho_{i}\sim 1$, we
have $z_{i}\lesssim 1$ (and therefore $|z_{e}|\ll 1$) associated with the  maximum growth
rate, and the ion kinetic effects, such as ion Landau damping and finite-ion-Larmor-radius
effects, significantly influence the growth rate and nonlinearly saturated wave amplitude. 
In Fig.\ref{fig0} we present the plot of $|z_{i}|$ versus $k_{\bot} \rho_{i}$ for different values of the relation of ion to electron temperatures. It displays, that only for  the ion temperature less than the electron temperature the hydrodynamic approximation
$|z_{i}|\gg 1$ is valid.

Numerical solution to Eq.(\ref{7}) is necessary because simplifying the dispersion
equation 
(\ref{7}) eliminates the ability to properly account for the ion kinetic effects.
The results 
of the numerical solution of Eq.(\ref{7}), which confirm the 
importance of the ion kinetic effects, are presented on Figs.\ref{fig1}--\ref{fig4}. 
Parameters considered pertinent to the conditions of SOL of 
Tokamaks: $\rho_{i}/L_{n} =0.002$, $v_{di}/v_{Ti}= - 0.002$. Also, we assume, that
$k_{x}
=k_{y}= k_{\bot}/\sqrt{2}$. In Fig.\ref{fig1}, we present the plots of normalized
frequency $
\omega/\omega_{ci}$, normalized growth rate $\gamma/\omega_{ci}$, $\left|z_{i}
\right|$ and $
\left|z_{e} \right|$ as a function $k_{\bot}\rho_{i}$ for $V'_{0}/\omega_{ci} =
0.001$, 
$T_{i}/T_{e} = 1.0$ and $\left( k_{z}\rho_{i}\right) ^{-1} = 1800$ for different
values of 
the parameter $\eta_{i}$. The main result of Fig.\ref{fig1} is that the maximum
growth rate 
is obtained in the region  $k_{\bot}\rho_{i}\sim1$, where $|z_{i}|\lesssim1$. It is
worth 
noting that for the case of negligible parallel-flow shear shown in Fig.\ref{fig1},
the 
effect of the ion temperature gradient, $\eta_{i}$, is pronounced.

The plots of normalized frequency $\omega/\omega_{ci}$, normalized growth rate $\gamma/
\omega_{ci}$, $\left|z_{i} \right|$ and $\left|z_{e} \right|$ as a function of
$\left(k_{z}\rho_{i}\right)^{-1}$ for parameters $\eta_{i} = 3.0$, $T_{i}/T_{e} = 1.0$ and
$k_{\bot}\rho_{i}=0.9$, for which the instability was predicted by Fig.\ref{fig1}, are
presented in ig.\ref{fig2} and Fig.\ref{fig3}. We find that the region of the maximum growth 
rate corresponds to $\left( k_{z}\rho_{i}\right) ^{-1}\sim2000$. It is interesting to
note, that we obtain $|z_{i}|\sim1$ for that region.

Fig.\ref{fig4} shows the normalized frequency $\omega/\omega_{ci}$, normalized
growth rate $\gamma/\omega_{ci}$, $\left|z_{i} \right|$ and $\left|z_{e} \right|$ as a 
function of $\left( V'_{0}/\omega_{ci}\right) ^{-1}$ for different values of $\eta_{i}$.  The 
parameters $k_{\bot}\rho_{i}\sim1$, $\left( k_{z}\rho_{i}\right) ^{-1}\sim2000$ were
used, at which the growth gate attains maximal value according to
Figs.\ref{fig1}--\ref{fig3}. We find that the instability requires the presence of the ion 
temperature gradient $\eta_{i}$ when the value of velocity shear is small, consistent with 
the interpretation of Fig.\ref{fig1}. Common to Figs.\ref{fig1}--\ref{fig4}, $|z_{i}|\sim1$ 
in the region of maximum growth rate.

It follows from Fig.\ref{fig4} that at small values of $\left(
V'_{0}/\omega_{ci}\right) 
^{-1}$, we have $|z_{i}|> 1$, independent of parameter $\eta_{i}$, 
and the D'Angelo instability\cite{Chibisov} develops. As $\left( V'_{0}/
\omega_{ci}\right) ^{-1}$ increases, and at small values of $\eta_{i}$ parameter,
the growth 
rate decreases and eventually becomes negative. At $\eta_{i}\geq 3$, the growth rate
is 
positive and no longer depends on the sign or magnitude of $\left(
V'_{0}/\omega_{ci}\right) 
^{-1}$.

\section{The kinetic, combined ion-temperature-gradient parallel-flow shear-driven
(ITG-SFD) instability}
\label{sec4}

Extending beyond the near-marginal-stability analysis of Rogister et
al.\cite{Rogister}, we consider 
the dispersion properties of the ITG-SFD instability for $|z_{i}|$ comparable to
unity, 
values at which the growth rate of ITG--SFD  instability is maximum. For these
values of $|
z_{i}|$, we take advantage of a Pade approximation for $W\left(z_{i}
\right)$ in the form
\begin{eqnarray}
&\displaystyle W\left(z_{i}\right) \approx
\frac{\sqrt{\pi}}{\sqrt{\pi}-2iz_{i}},\label{16}
\end{eqnarray}
and apply it to Eq.(\ref{8}). As a result, we obtain the simplest approximate
dispersion 
equation for the kinetic ITG--SFD instability for the $|z_{i}|\lesssim 1$ domain,
\begin{eqnarray}
&\displaystyle z_{i}^{2}A_{0i}\eta_{i}\left|\chi_{i}\right|\left(\pi-2 \right)
\nonumber \\  &\displaystyle
+z_{i}\left[\left( \pi -2\right) A_{0i}
\left(1-\frac{k_{y}V'_{0}}{k_{z}\omega_{ci}} \right)\right.
\nonumber \\  &\displaystyle
 \left. -2\left(1+\frac{T_{i}}{T_{e}}-A_{0i} \right)
-i\sqrt{\pi}\eta_{i}\left|\chi_{i}\right|A_{0i} \right]
\nonumber \\  &\displaystyle
 +\pi \left|\chi_{i}\right|
\left[A_{0i}\left(1-\frac{\eta_{i}}{2} \right)
-\eta_{i}k_{\bot}^{2}\rho_{i}^{2}\left(A_{0i}-A_{1i} \right)   \right]
\nonumber \\  &\displaystyle -
i\sqrt{\pi}\left(1+\frac{T_{i}}{T_{e}}-A_{0i}\frac{k_{y}V'_{0}}{k_{z}\omega_{ci}}
\right)=0.
\label{17}
\end{eqnarray}

\begin{figure}[htb]
\includegraphics[width=0.35\textwidth]{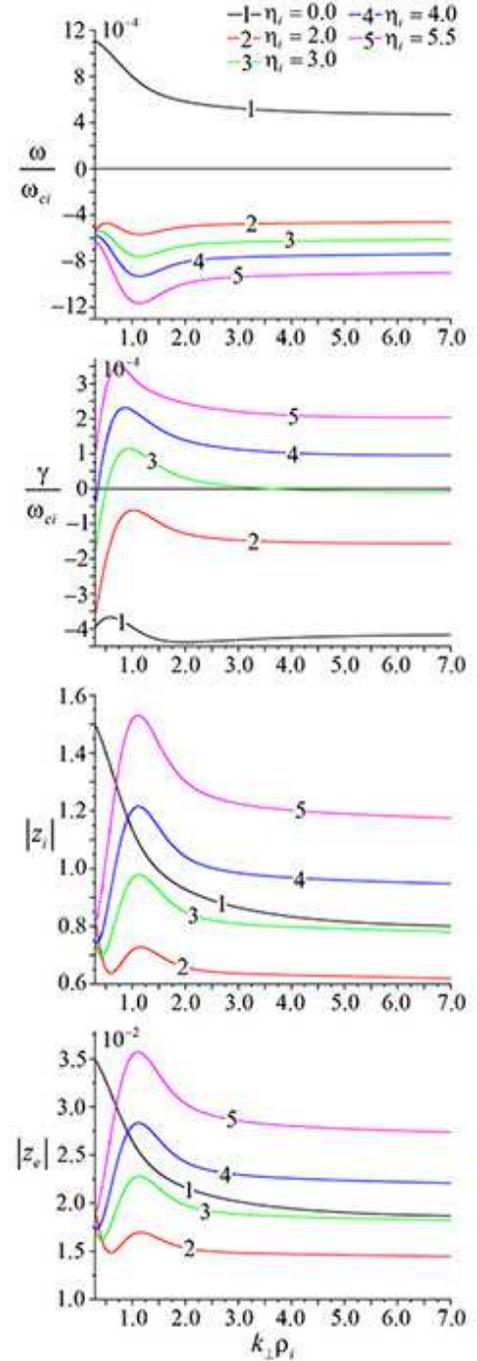}
\caption{\label{fig1} The normalized frequency $\omega/\omega_{ci}$, normalized
growth rate $\gamma/\omega_{ci}$, $\left|z_{i} \right|$ and $\left|z_{e} \right|$
versus $k_{\bot}\rho_{i}$ for $V'_{0}/\omega_{ci} = 0.001$, $T_{i}/T_{e} = 1.0$ 
and $\left( k_{z}\rho_{i}\right) ^{-1} = 1800$.}
\end{figure}

\begin{figure}[htb]
\includegraphics[width=0.35\textwidth]{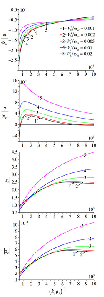}
\caption{\label{fig2} The normalized frequency $\omega/\omega_{ci}$, normalized
growth rate 
$\gamma/\omega_{ci}$,
$\left|z_{i} \right|$ and $\left|z_{e} \right|$
versus $k_{z}\rho_{i}$ for $\eta_{i} = 3.0$, $T_{i}/T_{e} = 1.0$ and
$k_{\bot}\rho_{i} = 
0.9$.}
\end{figure}

\begin{figure}[htb]
\includegraphics[width=0.35\textwidth]{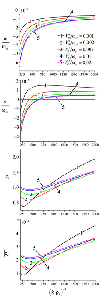}
\caption{\label{fig3} The normalized frequency $\omega/\omega_{ci}$, normalized
growth rate $\gamma/\omega_{ci}$, $\left|z_{i} \right|$ and $\left|z_{e} \right|$
versus $(k_{z}\rho_{i})^{-1}$ between 200 and 2000, for $\eta_{i} = 3.0$,
$T_{i}/T_{e} = 1.0$ 
and $k_{\bot}\rho_{i} = 0.9$.}
\end{figure}

\begin{figure}[htb]
\includegraphics[width=0.35\textwidth]{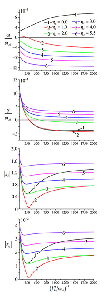}
\caption{\label{fig4} The normalized frequency $\omega/\omega_{ci}$, normalized
growth rate $\gamma/\omega_{ci}$, $\left|z_{i} \right|$ and $\left|z_{e} \right|$
versus $\left( V'_{0}/\omega_{ci}\right) ^{-1}$ for $k_{\bot}\rho_{i} = 0.9$,
$T_{i}/T_{e} = 1.0$ and $\left( k_{z}\rho_{i}\right) ^{-1} = 1800$.}
\end{figure}

This algebraic equation for $z_{i}$, in which all terms are assumed to be of the
same order 
of value, is not more complicated analytically than the dispersion equation for the
drift 
instabilities obtained in the opposite limit $|z_{i}| \gg 1$. It can be easily
solved with 
solution
\begin{eqnarray}
&\displaystyle \frac{\omega_{i1,2}}{\sqrt{2}k_{z}v_{Ti}}
=\frac{A_{0i}\pi\left(1-\frac{k_{y}V'_{0}}{k_{z}\omega_{ci}} \right)
-2\left(1+\frac{T_{i}}{T_{e}}-A_{0i}\frac{k_{y}V'_{0}}{k_{z}\omega_{ci}}  \right)  }
{2A_{0i}\eta_{i}\left|\chi_{i}\right|\left(\pi-2 \right)}
\nonumber \\  &\displaystyle
\pm\sqrt{\frac{r+x}{2} }, \label{18}
\\  &\displaystyle
 \frac{\gamma_{i1,2}}{\sqrt{2}k_{z}v_{Ti}}
=\frac{\sqrt{\pi}  }
{2\left(\pi-2 \right)}
\pm\sqrt{\frac{r-x}{2} },\label{19}
\end{eqnarray}
where $r=\sqrt{x^{2}+y^{2}}$ and
\begin{eqnarray}
&\displaystyle x =\left[\frac{A_{0i}\pi\left(1-\frac{k_{y}V'_{0}}{k_{z}\omega_{ci}}
\right)
-2\left(1+\frac{T_{i}}{T_{e}}-A_{0i}\frac{k_{y}V'_{0}}{k_{z}\omega_{ci}}  \right)  }
{2A_{0i}\eta_{i}\left|\chi_{i}\right|\left(\pi-2 \right)} \right]^{2}
\nonumber \\  &\displaystyle
-\frac{\pi}{4\left(\pi-2 \right)^{2}}
\nonumber \\  &\displaystyle
-\frac{\pi}{\left(\pi-2 \right)}
\left[\frac{2-\eta_{i}}{2\eta_{i}}-k_{\bot}^{2}\rho_{i}^{2}\left(1-\frac{A_{1i}}{A_{0i}}
\right)  \right], 
\label{20}\\
&\displaystyle  
y =-\frac{\sqrt{\pi}}{2\left(\pi-2 \right)}
\left[\frac{1  }
{2A_{0i}\eta_{i}\left|\chi_{i}\right|\left(\pi-2 \right)} \right.
\nonumber \\  &\displaystyle
\times\left( A_{0i}\pi\left(1-\frac{k_{y}V'_{0}}{k_{z}\omega_{ci}} \right)
-2\left(1+\frac{T_{i}}{T_{e}}-A_{0i}\frac{k_{y}V'_{0}}{k_{z}\omega_{ci}} 
\right)\right)
\nonumber \\  &\displaystyle
+\left. 2\frac{1+\frac{T_{i}}{T_{e}}-A_{0i}\frac{k_{y}V'_{0}}{k_{z}\omega_{ci}}}
{A_{0i}\eta_{i}\left|\chi_{i}\right|}  \right].\label{21}
\end{eqnarray}
The "quick" solution (\ref{18}) and (\ref{19}) with sign "+"  yields for the growth
rate  
satisfactory accuracy (within five percent relative error) compared to the numerical
solution 
of Eq.(\ref{7}) presented in Fig.\ref{fig1}, for $\eta_{i}\geq 2$ and any 
values of $V'_{0}/\omega_{ci}>0$ in the region where the growth rate attains maximum
value.

\begin{figure}[htb]
\includegraphics[width=0.35\textwidth]{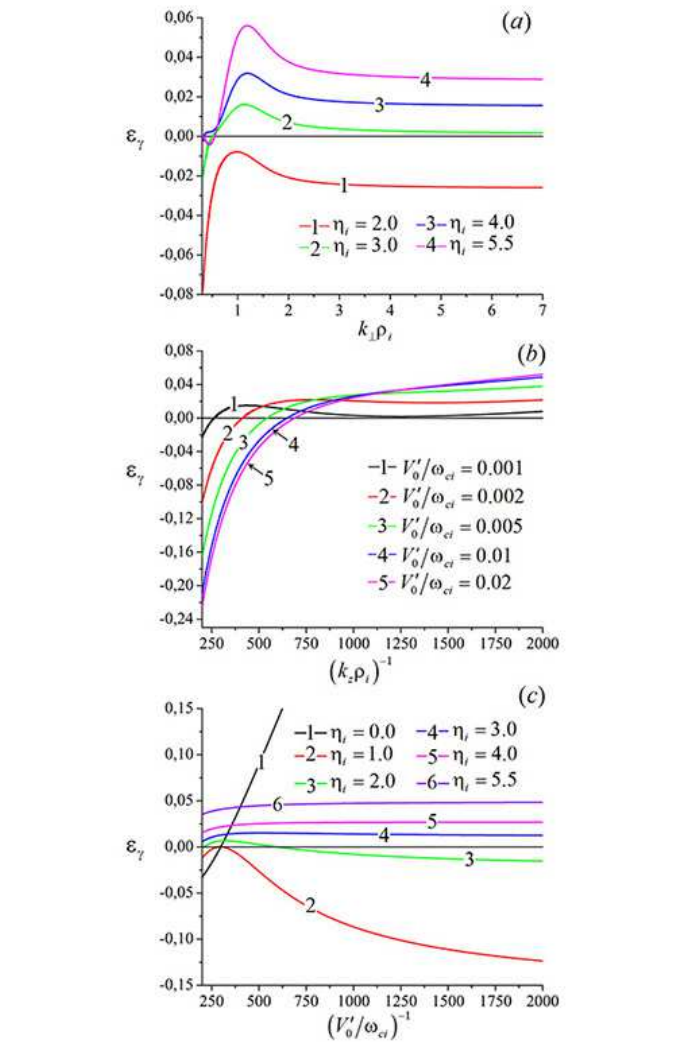}
\caption{\label{fig5} The relative error $\varepsilon_{\gamma}$ for the growth rate
$\gamma$: 
(a) versus $k_{\bot}\rho_{i}$ for $V'_{0}/\omega_{ci} = 0.001$, $T_{i}/T_{e} = 1.0$
and 
$k_{z}\rho_{i} = 0.00055$;
(b) versus $(k_{z}\rho_{i})^{-1}$ between 200 and 2000, for $\eta_{i} = 5.5$,
$T_{i}/T_{e} = 
1.0$ and $k_{\bot}\rho_{i} = 0.9$;
(c) versus $\left( V'_{0}/\omega_{ci}\right) ^{-1}$ for $k_{\bot}\rho_{i} = 0.9$,
$T_{i}/
T_{e} = 1.0$ and $k_{z}\rho_{i} = 0.00055$.}
\end{figure}

In the case of a plasma with homogeneous ion temperature, but with inhomogeneous
density,
i.e. for $\eta_{i} = 0$, this instability  continues to exist in the short
wavelength range 
as the kinetic D'Angelo instability\cite{Chibisov}, which is excited due to inverse
ion 
Landau damping. In Ref.\cite{Chibisov}, the wave real frequency and the growth rate
was 
obtained in the vicinity of the instability threshold. Eq.(\ref{17}) gives 
for $\eta_{i} = 0$ a simple expression for wave frequency and growth rate of the  ion 
kinetic D'Angelo instability for the range over which the growth rate is maximum:
\begin{eqnarray}
&\displaystyle \omega= \frac{\pi A_{0i}\left|k_{y}v_{di}\right|}{\left(\pi-2 \right)
A_{0i}\left(\frac{k_{y}V'_{0}}{k_{z}\omega_{ci}}-1 \right)
+2\left(1+\frac{T_{i}}{T_{e}}-
A_{0i} \right)  }
, \label{22}\\
&\displaystyle  \gamma=
\frac{\sqrt{2\pi}k_{z}v_{Ti}\left(A_{0i}\frac{k_{y}V'_{0}}{k_{z}\omega_{ci}}-\left( 1+
\frac{T_{i}}{T_{e}}
\right) \right)}
{\left(\pi-2 \right)
A_{0i}\left(\frac{k_{y}V'_{0}}{k_{z}\omega_{ci}} -1\right)
+2\left(1+\frac{T_{i}}{T_{e}} -
A_{0i}\right)}.\label{23}
\end{eqnarray}
It follows from Eq.(\ref{23}), that the instability exists for the $k_{z}\rho_{i}$
values 
below aa certain value   
\begin{eqnarray}
&\displaystyle k_{z}\rho_{i} <
\frac{V'_{0}}{\omega_{ci}}\frac{A_{0i}}{1+\frac{T_{i}}{T_{e}}}
k_{y}\rho_{i}.\label{24}
\end{eqnarray}
According to Fig.\ref{fig5}, the solution (\ref{19}) is valid to an accuracy of
better than
10\% of the relative error $\varepsilon_{\gamma}=\left(\gamma_{approx}
-\gamma_{exact}\right)/\gamma_{exact}$ over the wide intervals of the pertinent
parameters. 
In the case of the parallel-flow shear with homogeneous ion temperature
\cite{Chibisov} $(\eta_{i} = 0)$, the relative error between the approximate 
solution (\ref{23}) and exact numerical solution to Eq.(\ref{7}) (black line on Fig.
\ref{fig5}, case (c)) is less than 10\% only for large shear, i.e. $\left( V'_{0}/
\omega_{ci}\right) ^{-1}\leq 500$. The accuracy of the approximate solution (\ref{23}) 
improves with increasing value of $\eta_{i} \neq 0$. 

In Eqs.(\ref{18})--(\ref{19}), it was assumed, that $\eta_{i}\neq 0$. In the
different case 
of zero $\eta_{i}$,  zero $V'_{0}$ and $T_{i}\sim T_{e}$, Eqs.(\ref{18}) and
(\ref{19}) 
predict the  absence of the kinetic instability with $|z_{i}|\lesssim 1$,
\begin{eqnarray}
&\displaystyle \omega= \frac{\pi
A_{0i}\left|k_{y}v_{di}\right|}{2\left(1+\frac{T_{i}}{T_{e}} 
\right)-\pi A_{0i}}
, \label{25}\\
&\displaystyle  \frac{\gamma}{\sqrt{2}k_{z}v_{Ti}}=-
\frac{\sqrt{\pi}\left(1+\frac{T_{i}}{T_{e}} \right)}{2\left(1+\frac{T_{i}}{T_{e}}
\right)-\pi 
A_{0i}}.\label{26}
\end{eqnarray}

\section{DISCUSSIONS AND CONCLUSIONS}\label{sec5}

In this paper, we elucidated the thermal effects of ions having inhomogeneous
temperature profile on plasma stability in the presence of
parallel flow shear. On the basis of
the numerical solution of the general dispersion equation (\ref{7}) we confirmed
that the kinetic instability develops jointly with hydrodynamic D'Angelo instability
due to 
inverse ion Landau damping and has comparable real and imaginary values of frequency
at 
short wavelength over the interval having $k_{\perp} \rho_{i}$ of order unity.  

We find that increasing the ion-temperature gradient reduces the instability
threshold for the hydrodynamic D’Angelo instability and that a kinetic instability
arises from the coupled, reinforcing action of parallel-flow shear and
ion-temperature gradient. For the case of comparable electron and ion temperature,
we illustrate analytically the transition of the D'Angelo instability to the kinetic
instability as the shear parameter, the normalized wavenumber, and the ratio of
density-gradient and ion-temperature-gradient scale lengths are varied.

The approximate 
analytical solution of the dispersion equation, which uses a simple Pade approximation 
(\ref{16}) for the complex error function, is reported for the parameters associated
with the 
maximum growth rate. The approximate Pade solution of the dispersion equation was
compared 
with the numerical solution of the dispersion equation for the same plasma
conditions and for numerical  
parameters that may be pertinent to the conditions of the scrape-off-layer in
tokamaks. We 
find that the thermal motion of ions along the magnetic field may be important for
those plasma conditions and that improvement could be derived by incorporating the
parallel dynamics of ions along the magnetic field into the SOL codes. 
Although the dispersion equation  (\ref{7}) that accounted for the ion kinetic
effects is  the simplest one, it does not account for numerous effects such as the 
3-D inhomogeneity of the toroidal magnetic field, presence of the limiters or
divertors that lead to 
finite field-line length issues, or numerous other effects associated with the
processes taking place in a tokamak or stellarator. Because 
of unexplored dependencies of the shear parameter on the realistic factors present
in toroidal geometry, the presented plots provide only qualitative estimates for the
frequency and growth rate for the D’Angelo instability in the SOL-edge layer having
inhomogeneous ion temperature. For the less complicated geometry of space plasma,
the estimates may prove to be more accurate. In both cases, intuition for
interpreting instability and the behavior of unstable modes can be derived from the results presented here.

\begin{acknowledgments}
This work was funded by National R$\&$D Program through the National
Research Foundation of Korea(NRF) funded by the Ministry of Education, Science and
Technology (Grant No.2013005758). Coauthor MK gratefully acknowledges support from 
U.S. NSF grant NSF-PHYS-0613238.
\end{acknowledgments}

\end{document}